\documentclass[aps,pra,showpacs,twocolumn]{revtex4-1}
\usepackage{amssymb}
\usepackage{amsmath}
\usepackage{graphicx}
\usepackage{epsfig}
\usepackage{subfigure}
\usepackage{amsfonts}
\usepackage{color}
\usepackage{bm}
\usepackage{empheq}

\begin{document}

\title{Generalized phantom helix states in quantum spin graphs}
\author{C. H. Zhang}
\author{Y. B. Shi}
\author{Z. Song}
\email{songtc@nankai.edu.cn}

\begin{abstract}
In general, the summation of a set of sub-Hamiltonians cannot share a common
eigenstate of each one, only if it is an unentangled product state, such as
a phantom helix state in quantum spin system. Here we present a method,
referred to as the building block method (BBM), for constructing possible spin-$%
1/2$ XXZ Heisenberg lattice systems possessing phantom helix states. We
focus on two types of XXZ dimers as basic elements, with a non-Hermitian
parity-time ($\mathcal{PT}$) field and Hermitian Dzyaloshinskii-Moriya
interaction (DMI),  which share the same degenerate
eigenstates. Based on these two building blocks, one can construct a variety
of Heisenberg quantum spin systems, which support helix states with zero
energy. The underlying mechanism is the existence of a set of degenerate
eigenstates. Furthermore, we show that such systems act as quantum spin
graphs since they obey the  analogs of Kirchhoff's laws for sets of
spin helix states when the non-Hermitian $\mathcal{PT}$ fields cancel each
other out. In addition, the dynamic response of the helix states for three
types of perturbations is also investigated analytically and numerically.
Our findings provide a way to study quantum spin systems with irregular
geometries beyond the Bethe ansatz approach.
\end{abstract}

\affiliation{School of Physics, Nankai University, Tianjin 300071, China }
\maketitle

\section{Introduction}

\label{Introduction} The study of dynamical excitations becomes crucial to
the understanding of many exotic quantum phenomena in artificial quantum
materials. Recent advances in experimental capability \cite%
{kinoshita2006quantum,trotzky2012probing,gring2012relaxation,schreiber2015observation,smith2016many,kaufman2016quantum}
have stimulated the study of\ the nonequilibrium dynamics of quantum
many-body systems, which has emerged as a fundamental and attractive topic
in condensed-matter physics. In particular, as an excellent test bed for
quantum simulators in experiments \cite%
{zhang2017observation,bernien2017probing,barends2015digital,davis2020protecting,signoles2021glassy,trotzky2008time,gross2017quantum}%
, the atomic system allows us to verify the theoretical predictions on the
dynamics of quantum spin systems, which not only capture the properties of
many artificial systems, but also provide tractable theoretical examples for
understanding fundamental concepts in physics. A well-known example is the
spin-1/2 XXZ Heisenberg chain, which is integrable and therefore most of its
properties may be obtained exactly \cite{Mikeska2004one}. Recently, the
discovery of highly excited many-body eigenstates of the Heisenberg model,
referred to as Bethe phantom states, has received much attention from both
theoretical \cite%
{popkov2016obtaining,popkov2017solution,popkov2020exact,popkov2021phantom,MESPRB}
and experimental approaches \cite%
{jepsen2020spin,jepsen2021transverse,hild2014far,jepsen2022long}. However,
the well-known exact result is rare due to the limitation of the Bethe-Ansatz
technique, which usually requires the system to be one-dimensional and
uniform.

In this work, we focus on the extension of the obtained result in a uniform
system to those with irregular geometries. To this end, we propose a
building block method (BBM), for constructing possible spin-$1/2$ XXZ
Heisenberg lattice systems possessing phantom helix states. We present two
types of XXZ dimers as basic elements. One is two coupled spins with
a non-Hermitian parity-time ($\mathcal{PT}$) field, and the other is a Hermitian
one with a Dzyaloshinskii-Moriya interaction\ (DMI). Both  dimers share the
same degenerate eigenstates under resonant conditions. A variety of
Heisenberg quantum spin systems can be constructed based on these two
building blocks. These systems can be Hermitian or non-Hermitian, supporting
helix states with zero energy. The underlying mechanism is the existence of
a set of degenerate eigenstates, which are also eigenstates of the total spin
operator in the  $z$ direction. In general, the summation of a set of
sub-Hamiltonians cannot share a common eigenstate of each one, only if it
is an unentangled product state. The superposition of these degenerate
states can be a phantom helix state. When applying the scheme to construct a
Hermitian system, one should optimize the parameters to cancel\ out all
non-Hermitian fields. Such restrictions allow the system to act as quantum spin
graphs in the following context\textbf{.}\ For sets of spin helix states,
the corresponding spin currents and helical phases obey the analogs of
Kirchhoff's laws. In addition, the dynamic response of the helix states\ for
three types of perturbations is also investigated analytically and
numerically. Our findings provide a way to study quantum spin systems with
irregular geometries beyond the Bethe ansatz approach.

\begin{figure}[h]
\centering\includegraphics[width=0.45\textwidth]{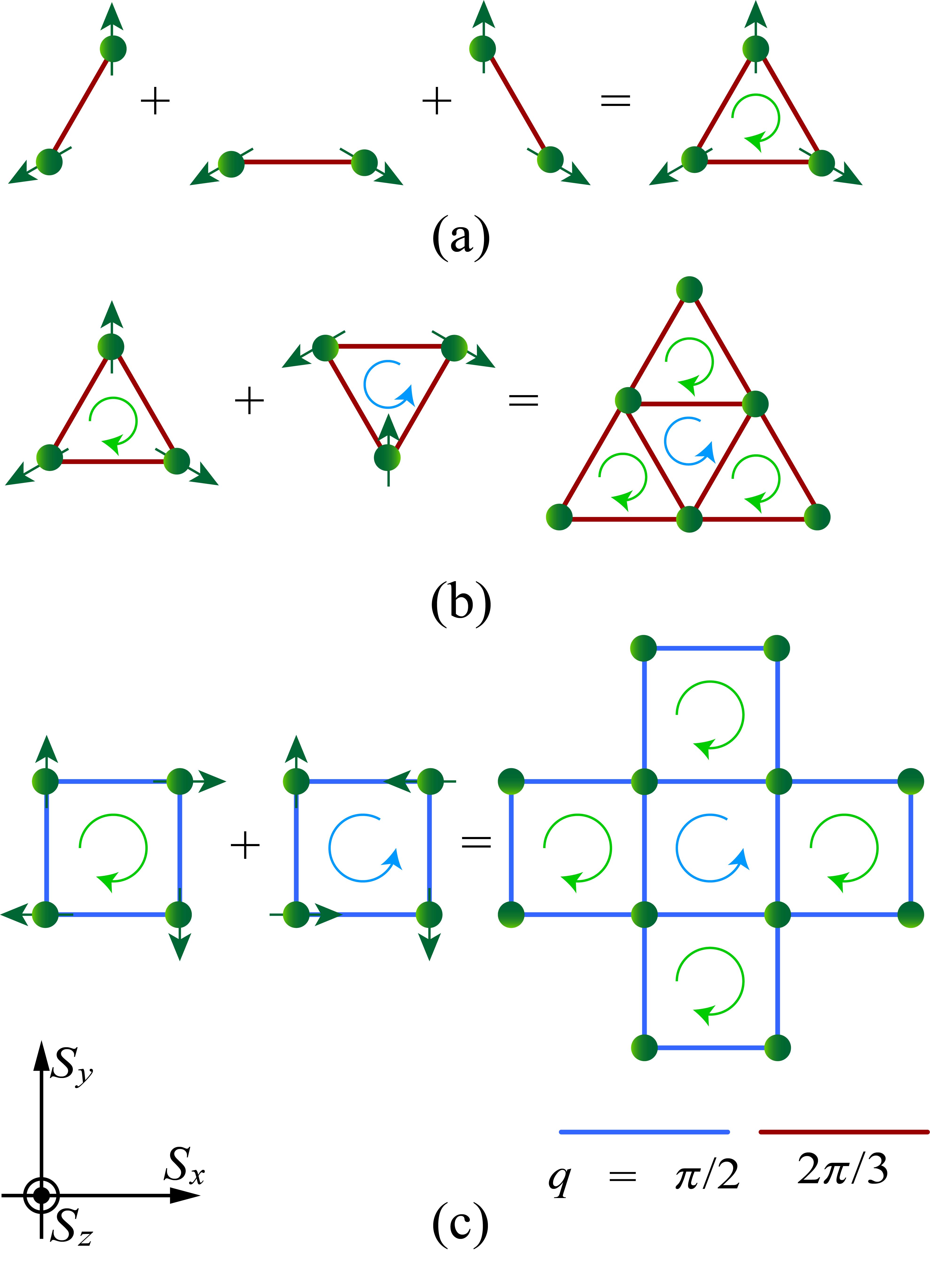} 
\caption{Schematic illustration of the building block method (BBM) via
simple systems. (a) Eigenstates of XXZ dimers with $q=2\protect\pi /3$, which
are product states of two spins, indicated by two vectors in the  $xy$ plane,
spin states. Three dimer eigenstates are taken as the configurations: each
pair of dimer eigenstates shares the same spin state, satisfying the conditions
of the BBM. Then, a spin helix eigenstate of a triangle XXZ model can be
constructed in the form of the right. (b) Two eigenstates of the XXZ triangle
with $q=\pm 2\protect\pi /3$, indicated by two opposite curved arrows,
respectively. Based on the two types of triangles as building blocks, a
triangular lattice fragment can be constructed, which possesses a spin helix
eigenstate. Note that two neighboring triangles have opposite helicities. (c)
The same construction for the square lattice fragment. All obtained systems are
Hermitian.} \label{fig1}
\end{figure}

The rest of this paper is organized as follows: In Sec.~\ref{Hamiltonian and
building block method}, we introduce the model Hamiltonian and the building
block method. Based on the exact solution, in Sec.~\ref{Two types of dimers}
we present two types of dimer systems as building blocks. In Sec.~\ref%
{Kirchhoff's laws in spin graphs}, we apply the method to Hermitian quantum
spin systems and elucidate the analog of Kirchhoff's laws for constructed
spin helix states. The dynamic response of the helix states\ for three types
of perturbations is presented in Sec.~\ref{Quenching dynamics}. Sec.~\ref%
{Summary} concludes this paper.

\section{Hamiltonian and building block method}

\label{Hamiltonian and building block method} We start with a general result
for the eigenstate of a quantum spin system, based on which a building block
method can be developed. This method can be extended to more generalized
fermion and boson systems without any particular restrictions on the
dimensionality and geometry.

\textit{Theorem on quantum spin system}. We consider two Hamiltonians $%
H_{1}(a,c)$ and $H_{2}(b,c)$ on two sets of spins, which share the same
subset of spins. Here $a$, $b$, and $c$ label three sets of spins, for
instance, $a\subset \left\{ s_{j}\text{, }j\in \left[ 1,M\right] \right\} $, 
$b\subset \{s_{j}$, $j\in \lbrack M+1$, $M+N]\}$, and $c\subset \{s_{j}$, $%
j\in \lbrack M+N+1$, $M+N+L]\}$, where $M$, $N$, and $L$ are the numbers of
spins of three sets, respectively. If two Hamiltonians have eigenstates of
product form 
\begin{equation}
H_{1}(a,c)\left\vert \psi _{a}(a)\right\rangle \left\vert \psi
_{c}(c)\right\rangle =E_{1}\left\vert \psi _{a}(a)\right\rangle \left\vert
\psi _{c}(c)\right\rangle ,  \label{ac}
\end{equation}%
and%
\begin{equation}
H_{2}(b,c)\left\vert \psi _{b}(b)\right\rangle \left\vert \psi
_{c}(c)\right\rangle =E_{2}\left\vert \psi _{b}(b)\right\rangle \left\vert
\psi _{c}(c)\right\rangle ,  \label{bc}
\end{equation}%
then one of the eigenstates of $H_{1}+H_{2}$ can be obtained as the form $%
\left\vert \psi _{abc}\right\rangle =\left\vert \psi _{a}(a)\right\rangle
\left\vert \psi _{b}(b)\right\rangle \left\vert \psi _{c}(c)\right\rangle $,
satisfying \ 
\begin{equation}
\left( H_{1}+H_{2}\right) \left\vert \psi _{abc}\right\rangle =\left(
E_{1}+E_{2}\right) \left\vert \psi _{abc}\right\rangle .  \label{abc}
\end{equation}

The proof of the theorem is straightforward. Extending Eqs. (\ref{ac})
and (\ref{bc}) to the whole Hilbert space, i.e., 
\begin{equation}
H_{1}\left\vert \psi _{a}\right\rangle \left\vert \psi _{c}\right\rangle
\left\vert \psi _{b}\right\rangle =E_{1}\left\vert \psi _{a}\right\rangle
\left\vert \psi _{c}\right\rangle \left\vert \psi _{b}\right\rangle ,
\end{equation}%
and%
\begin{equation}
H_{2}\left\vert \psi _{b}\right\rangle \left\vert \psi _{c}\right\rangle
\left\vert \psi _{a}\right\rangle =E_{2}\left\vert \psi _{b}\right\rangle
\left\vert \psi _{c}\right\rangle \left\vert \psi _{a}\right\rangle ,
\end{equation}%
the sum of two equations results in the conclusion (\ref{abc}).

This conclusion can be extended to the case with multiple sub-Hamiltonians.
A trivial illustrative example is a spin-$1/2$ isotropic Heisenberg model on
an arbitrary lattice, with the Hamiltonian%
\begin{equation}
H_{\text{iso}}=\sum_{i>j}J_{ij}\left( \mathbf{s}_{i}\cdot \mathbf{s}%
_{j}-1/4\right) ,
\end{equation}%
where the coupling strength $\left\{ J_{ij}\right\} $\ is a set of arbitrary
numbers\ and the spin operator $\mathbf{s}_{j}=\left(
s_{j}^{x},s_{j}^{y},s_{j}^{z}\right) $. The Hamiltonian can be rewritten  as the
sum of a set of Heisenberg dimers $\left\{ H_{i,j}\right\} $, with%
\begin{equation}
H_{i,j}=J_{ij}\left( \mathbf{s}_{i}\cdot \mathbf{s}_{j}-1/4\right) ,
\end{equation}%
which has a zero-energy ferromagnetic eigenstate in an arbitrary $\mathbf{n}$
direction 
\begin{equation}
\left\vert \text{FM}\right\rangle _{ij}=\left\vert \mathbf{n}\right\rangle
_{i}\left\vert \mathbf{n}\right\rangle _{j},
\end{equation}%
with $\left( \mathbf{s}_{i}\cdot \mathbf{n}\right) \left\vert \mathbf{n}%
\right\rangle _{i}=\frac{1}{2}\left\vert \mathbf{n}\right\rangle _{i}$ and $%
\left( \mathbf{s}_{j}\cdot \mathbf{n}\right) \left\vert \mathbf{n}%
\right\rangle _{j}=\frac{1}{2}\left\vert \mathbf{n}\right\rangle _{j}$.
Obviously, state%
\begin{equation}
\left\vert \text{FM}\right\rangle =\prod_{j}\left\vert \mathbf{n}%
\right\rangle _{j},
\end{equation}%
is a zero-energy ferromagnetic eigenstate in an arbitrary $\mathbf{n}$
direction of the total Hamiltonian $H_{\text{iso}}$. This is a direct
demonstration of the theorem. It is expected that this theorem can be
applied to and get insight into other nontrivial quantum systems. The goal of
this work focuses on the spin helix state in the XXZ Heisenberg model on a
lattice beyond a one-dimensional and uniform system.

\section{Two types of dimers}

\label{Two types of dimers}

We see that the eigenstate $\left\vert \psi _{abc}\right\rangle $ of the
entire Hamiltonian is a product state. Conversely, such a state should also
be the eigenstate of each sub-Hamiltonian.\ Accordingly, if an eigenstate of
an XXZ model is a spin helix state, it should be the eigenstate of a
sub-Hamiltonian on a dimer. Our strategy is taking the XXZ dimer\ as the
starting point, which is regarded as a building block to construct the target
Hamiltonian.

We consider two types of XXZ dimers, with the Hamiltonians%
\begin{eqnarray}
\mathcal{H}_{i,j}(q) &=&\left( s_{i}^{x}s_{j}^{x}+s_{i}^{y}s_{j}^{y}\right)
+\cos q\left( s_{i}^{z}s_{j}^{z}-1/4\right)  \notag \\
&&+\frac{i}{2}\sin q\left( s_{i}^{z}-s_{j}^{z}\right) ,
\end{eqnarray}%
and%
\begin{eqnarray}
H_{i,j}(q) &=&\cos q\left( s_{i}^{x}s_{j}^{x}+s_{i}^{y}s_{j}^{y}\right)
+\left( s_{i}^{z}s_{j}^{z}-1/4\right)  \notag \\
&&+\sin q\left( s_{i}^{x}s_{j}^{y}-s_{i}^{y}s_{j}^{x}\right) ,
\end{eqnarray}%
respectively, where $q$\ is an arbitrary real number. Here $\mathcal{H}%
_{i,j} $\ is a XXZ dimer with a non-Hermitian parity-time ($\mathcal{PT}$)
field in $z$ direction, while $H_{i,j}$\ is one with a Hermitian
Dzyaloshinskii-Moriya interaction\ (DMI). Straightforward derivation shows
that both $\mathcal{H}_{i,j}(q)$\ and $H_{i,j}(q)$\ share the same
eigenstates%
\begin{eqnarray}
\left\vert \phi _{1}\right\rangle &=&\left\vert \uparrow \right\rangle
_{i}\left\vert \uparrow \right\rangle _{j},\left\vert \phi _{2}\right\rangle
=\left\vert \downarrow \right\rangle _{i}\left\vert \downarrow \right\rangle
_{j}, \\
\left\vert \phi _{3}\right\rangle &=&e^{iq/2}\left\vert \uparrow
\right\rangle _{i}\left\vert \downarrow \right\rangle
_{j}+e^{-iq/2}\left\vert \downarrow \right\rangle _{i}\left\vert \uparrow
\right\rangle _{j},
\end{eqnarray}%
with zero energy. We note that $H_{12}=\mathcal{H}_{12}$\ reduces to an 
isotropic Heisenberg dimer when $q=0$.

In general, a zero-energy eigenstate can be constructed as the product form%
\begin{eqnarray}
&&\left\vert \psi _{i}(\theta ,\Omega -q/2)\right\rangle \left\vert \psi
_{j}(\theta ,\Omega +q/2)\right\rangle =\cos ^{2}\left( \theta /2\right)
\left\vert \phi _{1}\right\rangle  \notag \\
&&+e^{i2\Omega }\sin ^{2}\left( \theta /2\right) \left\vert \phi
_{2}\right\rangle +e^{i\Omega }\sin \theta \left\vert \phi _{3}\right\rangle
,
\end{eqnarray}%
for arbitrary\ $\Omega $\ and $\theta $, with%
\begin{equation}
\left\vert \psi _{j}(\theta ,Q)\right\rangle =\cos \left( \theta /2\right)
\left\vert \uparrow \right\rangle _{j}+e^{iQ}\sin \left( \theta /2\right)
\left\vert \downarrow \right\rangle _{j}.
\end{equation}%
It indicates that a common eigenstate of dimers $\mathcal{H}_{i,j}(q)$\ and $%
H_{i,j}(q)$\ can be a segment of the  helix state. According to the above theorem
the zero energy eigenstate of a Hamiltonian 
\begin{equation}
H=\sum_{i<j}\left( \alpha _{ij}H_{i,j}+\beta _{ij}\mathcal{H}_{i,j}\right) ,
\end{equation}%
can be constructed in the form $\prod_{j}\left\vert \psi _{j}(\theta
,Q_{j})\right\rangle $ for an  arbitrary set of numbers $\left\{ \alpha
_{ij},\beta _{ij}\right\} $ under certain conditions for the geometry of the
lattice. The choose of the phase $\Omega $\ in $\left\vert \psi
_{j}\right\rangle $\ is crucial for a given Hamiltonian under some
constraints on the distribution of the value of $q_{ij}$ in each dimer ($%
i,j $), resulting in the helix state%
\begin{eqnarray}
&&\left\vert \Phi \right\rangle =\prod _{j=1}^{N}\left\vert \psi _{j}(\theta ,\Lambda (j))\right\rangle \notag\\
&&=\prod_{j=1}^{N}\left[ \cos \left( \theta
/2\right) \left\vert \uparrow \right\rangle _{j}+e^{i\Lambda (j)}\sin \left(
\theta /2\right) \left\vert \downarrow \right\rangle _{j}\right] ,
\label{phi}
\end{eqnarray}%
where $\Lambda (j)-\Lambda (i)=q_{ij}$. Here we consider two Hamiltonians on
an $N$-site ring,%
\begin{equation}
\mathcal{H}_{\text{\textrm{ring}}}=\sum_{j=1}^{N}\mathcal{H}_{j,j+1},H_{%
\text{\textrm{ring}}}=\sum_{j=1}^{N}H_{j,j+1},
\end{equation}%
with $q=2\pi n/N$ ($n\in \left[ 1,N\right] $). Both $\mathcal{H}_{\text{ 
\textrm{ring}}}$\ and $H_{\text{\textrm{ring}}}$\ are uniform Hermitian
systems, and have the same spin helix state%
\begin{equation}
\prod_{j=1}^{N}\left[ \cos \left( \theta /2\right) \left\vert \uparrow
\right\rangle _{j}+e^{iqj}\sin \left( \theta /2\right) \left\vert \downarrow
\right\rangle _{j}\right] ,
\end{equation}%
which has been proposed in Refs. \cite{MESPRB} and \cite{SYBandZSPhantom}.

\begin{figure}[h]
\centering\includegraphics[width=0.45\textwidth]{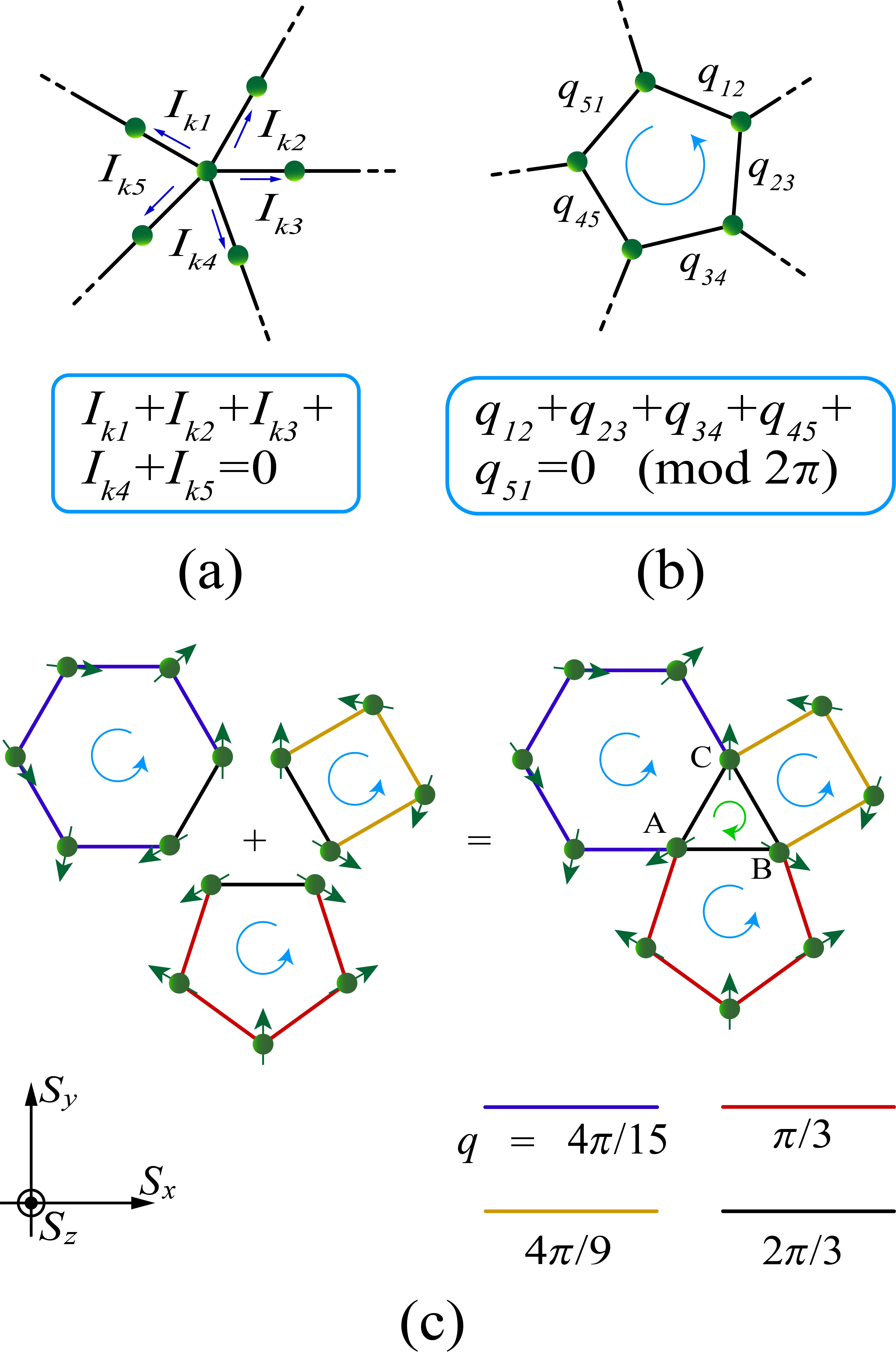} 
\caption{Schematic illustrations of Kirchhoff's laws in a spin graph\ by two
examples (a) and (b), a star-shaped graph and simple loop. (a) The spin
current entering any site is equal to the current leaving that site.\ The
five currents are indicated in the graph and the formula. (b) The sum of all
the $q$ values of dimers around a loop is equal to zero (mode $2\protect\pi $). The five q values are indicated in the graph and the formula. (c)
Schematic for a fragment of the lattice with mixed $q$ values, which
consists of one quadrilateral, pentagon, and hexagon lattice. The distibution
of $q$ values\ in dimers is indicated by different colors. Here, sites $A$, $B$, and $C$ are three vertices of the equilateral triangle. The entanglement
entropy and current about site $A$ are plotted in {Fig. \protect\ref{fig3}.}}
\label{fig2}
\end{figure}

\section{Kirchhoff's laws in spin graphs}

\label{Kirchhoff's laws in spin graphs}

Now we focus on the constraint on the lattice systems which are constructed
based on the two types of dimers and have a spin helix state. Any lattice
system can be regarded as a spin graph  consisting of nodes (spins) and
edges (dimers). Based on the analysis above, an XXZ Hamiltonian has
a zero-energy spin helix eigenstate, if the corresponding graph satisfies the
condition that for any loops we always have%
\begin{equation}
\sum_{\text{\textrm{loop}}}q_{ij}=0,(\text{\textrm{mod }}2\pi ).
\end{equation}%
One can imagine that the variety of such systems is enormous, including
Hermitian and non-Hermitian ones.

In the following, we restrict our study to Hermitian systems but
consisting of non-Hermitian dimers only. The corresponding Hamiltonian has
the form 
\begin{equation}
H_{\mathrm{gra}}=\ \sum_{i<j}R_{ij}\mathcal{H}_{i,j},
\end{equation}%
where $\left\{ R_{ij}\right\} $\ should be taken to meet $H_{\mathrm{gra}%
}=\left( H_{\mathrm{gra}}\right) ^{\dag }$ in this situation. The
Hamiltonians of such systems are independent of the signs of $q_{ij}$,
allowing the solution of bidirectional helix states. In order to meet the
condition of the Hermiticity of the local Hamiltonian $h_{k}=%
\sum_{j=1}^{n}R_{kj}\mathcal{H}_{k,j}$ around any node at the $k$ site, which
connects to $n$ dimers $\{\mathcal{H}_{k,j},j\in \left[ 1,n\right] \}$, the
cancellation of the imaginary fields requires%
\begin{equation}
\sum_{j=1}^{n}R_{kj}\sin q_{kj}=0.
\end{equation}%
Accordingly, the eigenstate of the local Hamiltonian $h_{k}$\ is%
\begin{eqnarray}
&&\left[ \cos \left( \theta /2\right) \left\vert \uparrow \right\rangle
_{k}+e^{iQ}\sin \left( \theta /2\right) \left\vert \downarrow \right\rangle
_{k}\right] \times  \notag \\
&&\prod_{j=1}^{n}\left[ \cos \left( \theta /2\right) \left\vert \uparrow
\right\rangle _{j}+e^{iQ}e^{iq_{kj}}\sin \left( \theta /2\right) \left\vert
\downarrow \right\rangle _{j}\right] .
\end{eqnarray}

Defining the spin current operator across the dimer $R_{kj}\mathcal{H}_{k,j}$
\begin{equation}
\hat{\jmath}_{kj}=4iR_{kj}\left( \sigma _{k}^{-}\sigma _{j}^{+}-\sigma
_{j}^{-}\sigma _{k}^{+}\right) ,
\end{equation}%
the current for the helix state $\left\vert \psi _{k}(\theta ,\Omega
-q/2)\right\rangle $ $\times \left\vert \psi _{j}(\theta ,\Omega
+q/2)\right\rangle $ is $-4R_{kj}\sin \theta \sin q$. Then the total current
flowing out of the node is%
\begin{eqnarray}
J_{k} &=&\sum_{j=1}^{n}I_{kj}=\sum_{j=1}^{n}\left\langle \hat{\jmath}%
_{kj}\right\rangle  \notag \\
&=&\sum_{j=1}^{n}-4R_{kj}\sin \theta \sin q_{kj}=0,
\end{eqnarray}%
for the spin helix states, {where }$\left\langle ...\right\rangle ${\
denotes the average of the helix state.} Together with the equation $\sum_{%
\text{\textrm{\ loop}}}q_{ij}=0,($\textrm{mod }$2\pi )$, these from the
analogues of Kirchhoff's laws.

We would like to point out that one of the main conclusions is that the spin
helix states do exist in the quantum XXZ model on irregular lattices with a
nonuniform $q${\ distribution }$\left\{ q_{ij}\right\} ${. Fig. \ref{fig2}%
(c) provides an example to demonstrate this point by a lattice fragment with
mixed }$q${\ values. The corresponding XXZ Hamiltonian consists of three
Hermitian sub-Hamiltonians, on quadrilateral, pentagon, and hexagon
lattices, respectively. The distribution }$\left\{ q_{ij}\right\} ${\ of }$q$%
{\ value\ is labeled in the figure and obviously, obeys the Kirchhoff's laws.}

\begin{figure}[h]
\centering\includegraphics[width=0.5\textwidth]{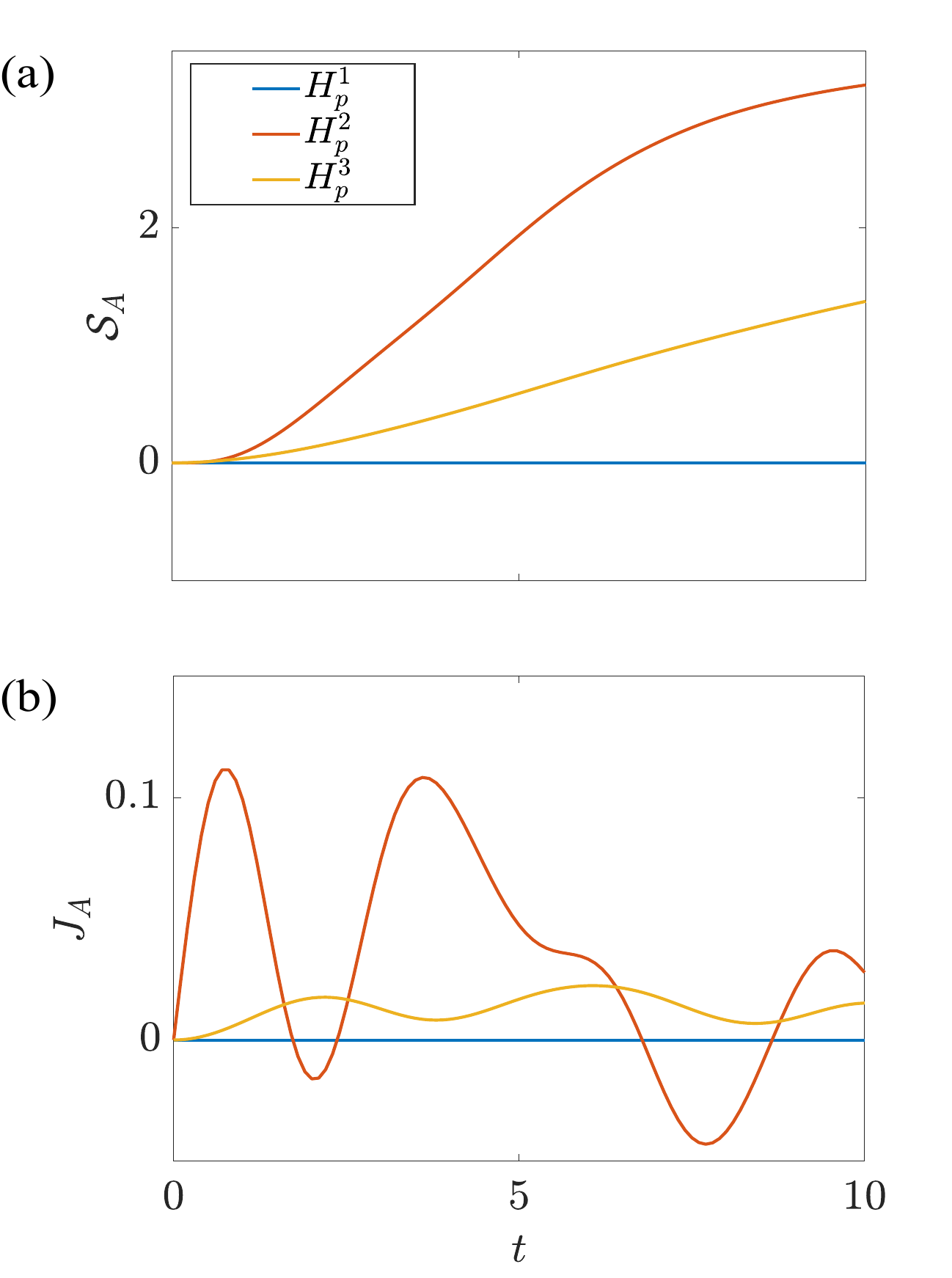} 
\caption{Plots of the entanglement entropy and current obtained from the
system in Fig. \protect\ref{fig2}(c) with different perturbation terms $H_{\mathrm{p}}^{1,2,3}$, obtained by exact diagonalization. (a) The
entanglement entropy is obtained from Eq. \protect\ref{Sa}\ for the time
evolution of the initial helix state, where the subsystem $A$ consists of the left six spins of the graph in the right panel of Fig. \protect\ref{fig2}(c). (b) The
current is obtained from Eq. \protect\ref{Jk}\ for the site $A$ ($k=A$). The
local field in $H_{\mathrm{p}}^{2}$\ is applied at site $A$ ,and other\
parameters in $H_{\mathrm{p}}^{1,2,3}$\ are $h=1$ and $\Delta q=0.05\protect\pi $. As expected, we find that\ $\mathcal{S}_{A}=J_{k}=0$\ for $H_{\mathrm{p}}^{1}$, while positive $\mathcal{S}_{A}$ and oscillating $J_{A}$ indicate
that $H_{\mathrm{p}}^{2,3}$ induces the deviation from the helix state.} %
\label{fig3}
\end{figure}

\section{Quenching dynamics}

\label{Quenching dynamics}

The eigenstates we obtained for a constructed Hamiltonian are stationary
helix states since the orientation for any spins is constant in time. As we
have seen,  the existence of such a spin helix solution is based on the
resonance condition of $\left\{ q_{ij}\right\} $\ and $\left\{
R_{ij}\right\} $\ in a given Hamiltonian. When the value of $q_{ij}$\ in a
dimer changes, the solution is invalid. On the other hand, it is shown that
a helix state in a uniform ring system becomes a dynamic helix state \cite%
{ZGandZSnonintegrable}.This arise a question: What happens for a generalized
helix state by quenching perturbations? In this section, we consider three
types of perturbation terms%
\begin{equation}
H_{\mathrm{p}}^{1}=h\sum_{j}s_{j}^{z},H_{\mathrm{p}}^{2}=hs_{l}^{z},
\end{equation}%
{and}%
\begin{equation}
H_{\mathrm{p}}^{3}=H\left( q_{ij}+\Delta q\right) -H\left( q_{ij}\right) ,
\end{equation}%
{arising from a uniform field, {a local field at the }}${{l}}${th site in the 
}${\mathbf{z}}${{\ direction,} }{and\ the deviation }$\Delta q${\ from the
resonant }$\left\{ q_{ij}\right\} ${\ value for each dimer, respectively.
For a system with uniform }$q${, }$H_{\mathrm{p}}^{3}${\ can be regarded as
a global shift in the Ising term.} We will focus on the effect of the
perturbations on the helix state in two aspects: the spin current and the
entanglement.

In the case with $H_{\mathrm{p}}^{1}$, the Hermiticity of $\mathcal{H}_{%
\mathrm{gra}}$\ ensures that it is a standard XXZ Heisenberg model and then
obeys

\begin{equation}
\left[ s^{z},\mathcal{H}_{\mathrm{gra}}\right] =0,
\end{equation}%
with $s^{z}=\sum_{j=1}^{N}s_{j}^{z}$. However, a helix state is not $s^{z}$\
conservative, although $H_{\mathrm{p}}^{1}$\ shares common eigenstates with $%
\mathcal{H}_{\mathrm{gra}}$. In addition, we can introduce collective
spin operators%
\begin{equation}
\tau ^{+}=\left( \tau ^{-}\right) ^{\dag }=\sum_{j=1}^{N}e^{i\Lambda
(j)}s_{j}^{+},
\end{equation}%
which satisfy the Lie algebra%
\begin{equation}
\left[ \tau ^{+},\tau ^{-}\right] =2s^{z},\left[ \tau ^{\pm },s^{z}\right]
=\mp 2\tau ^{\pm }.
\end{equation}%
Notably, the spin helix state we have defined in Eq.(\protect\ref{phi}) can be expressed in the form%
\begin{equation}
\left\vert \Phi \right\rangle =\sum_{n}d_{n}\left\vert \psi
_{n}\right\rangle ,
\end{equation}%
where the coefficient%
\begin{equation}
d_{n}=\sqrt{C_{N}^{n}}\sin ^{\left( N-n\right) }\left( \theta /2\right) \cos
^{\left( n\right) }\left( \theta /2\right) ,
\end{equation}%
and 
\begin{equation}
\left\vert \psi _{n}\right\rangle =\frac{1}{\Omega _{n}}\left( \tau
^{+}\right) ^{n}\prod_{j=1}^{N}\left\vert \downarrow \right\rangle _{j},
\end{equation}%
with the normalization factor $\Omega _{n}=\left( n!\right) \sqrt{C_{N}^{n}}$%
. Here the set of states $\{\left\vert \psi _{n}\right\rangle $, $n\subset %
\left[ 0,N\right] \}$\ are common eigenstates of $\mathcal{H}_{\mathrm{gra}}$%
\ and $H_{\mathrm{p}}^{1}$, i.e.,%
\begin{equation}
\mathcal{H}_{\mathrm{gra}}\left\vert \psi _{n}\right\rangle =0,H_{\mathrm{p}%
}^{1}\left\vert \psi _{n}\right\rangle =\frac{2n-N}{2}h\left\vert \psi
_{n}\right\rangle .
\end{equation}

When a uniform field $hs^{z}$\ is applied, the spin helix state $\left\vert
\Phi \right\rangle $\ is no longer the eigenstate of $\mathcal{H}_{\mathrm{gra}%
}+hs^{z}$, but its evolved state is still the eigenstate of $\mathcal{H}_{%
\mathrm{gra}}$. Considering $\left\vert \Phi \right\rangle $ as the initial
state at $t=0$, the evolved state is%
\begin{eqnarray}
&&e^{-i\left( \mathcal{H}_{\mathrm{gra}}+hs^{z}\right) t}\left\vert \Phi
\right\rangle =e^{i\frac{Nht}{2}}\sum_{n}e^{-inht}d_{n}\left\vert \psi
_{n}\right\rangle=e^{i\frac{Nht}{2}}   \notag \\
&& \prod_{j=1}^{N}[\cos \left( \theta /2\right) \left\vert \uparrow
\right\rangle _{j}+e^{-iht}e^{i\Lambda (j)}\sin \left( \theta /2\right)
\left\vert \downarrow \right\rangle _{j}],
\end{eqnarray}%
which is still a helix state and exhibits periodic dynamics. In this sense,
a dynamic helix state still obeys Kirchhoff's laws. {In the case of }$H_{%
\mathrm{p}}^{2}${, the perturbation term is not commutative with }$H_{%
\mathrm{gra}}${, and the evolved state should  not be a helix state.}

Now we turn to estimate the effect of a deviation $H_{\mathrm{p}}^{3}$ from $%
\mathcal{H}_{\mathrm{gra}}$ on a helix state in the framework of
perturbation theory. {Here, we consider the simplest case with a uniform }$q${\
distribution, }$q_{ij}=q${. Then, the perturbation term reduces to }%
\begin{equation}
H_{\mathrm{p}}^{3}=-\Delta q\sin q\sum_{<i,j>}s_{i}^{z}s_{j}^{z}\mathbf{.}
\end{equation}%
The matrix representation of the Hamiltonian $\mathcal{H}_{\mathrm{gra}}+H_{%
\mathrm{p}}^{3}$, in the subspace $\left\{ \left\vert \psi _{n}\right\rangle
\right\} $\ is a $\left( N+1\right) \times \left( N+1\right) $ matrix $M$
with nonzero matrix elements

\begin{equation}
M_{m,n}=-\Delta qN_{b}\sin q(\frac{1}{4}+\frac{n^{2}-Nn}{N^{2}-N})\delta
_{mn},
\end{equation}%
for $m$, $n=\left[ 1,N+1\right] $ ,where $N_{b}=\sum_{<i,j>}1$ is the total
number of dimers. It is obvious that for the initial helix state $%
\left\vert \Phi \right\rangle $ the {effective driven Hamiltonian is}$\ $%
\begin{equation}
H_{\mathrm{eff}}=H_{\mathrm{gra}}-\Delta qN_{b}\sin q[\frac{(s^{z})^{2}-N}{%
4N(N-1)}],
\end{equation}%
by ignoring a constant under the small $\Delta q$ approximation. The
relation $\left[ s^{z},H_{\mathrm{gra}}\right] =0${\ allows us to express
the evolved state in the form%
\begin{equation}
e^{-iH_{\mathrm{eff}}t}\left\vert \Phi \right\rangle =\sum_{n}e^{i\pi
(n^{2}-Nn)t/\tau }d_{n}\left\vert \psi _{n}\right\rangle ,
\end{equation}%
by ignoring an overall phase $e^{i\frac{\Delta qN_{b}\sin q}{4}t}$, which
exhibits periodic dynamics with period $\tau =\pi \frac{N(N-1)}{\Delta
qN_{b}\sin q}$ when $N$ is odd and $2\tau $ when $N$ is even. However, it is
no longer a helix state during this period, due to the nonlinear term $n^{2}$%
\ in the phase, and probably does not obey Kirchhoff's laws. In this sense, $%
H_{\mathrm{p}}^{3}$ may destroy the product state, resulting in an increase
in entanglement.}

Now we demonstrate and verify our predictions by numerical simulation on the
dynamic processes. We investigate the dynamic response of an initial helix
state on the perturbation $H_{\mathrm{p}}^{1,2,3}$. We measure the quenching
process by two quantities, the entanglement entropy and current, extracted from
the evolved state\ 
\begin{equation}
\left\vert \Psi (t)\right\rangle =e^{-i(\mathcal{H}_{\mathrm{gra}}+H_{%
\mathrm{p}}^{1,2,3})t}\left\vert \Phi \right\rangle ,
\end{equation}
{where the distribution }$\left\{ q_{ij}\right\} ${\ is not uniform in }$H_{%
\mathrm{p}}^{3}$ without loss of generality. For a given pure state $%
\left\vert \psi \right\rangle _{AB}$\ describing two sublattices $A$ and $B$%
, the entanglement entropy is defined as \cite{entanglement}%
\begin{equation}
\mathcal{S}_{A}=-\mathrm{Tr}[(\rho _{A}\log \rho _{A})],  \label{Sa}
\end{equation}%
where $\rho _{A}=\mathrm{Tr}_{B}(\left\vert \psi \right\rangle
_{AB}\left\langle \psi \right\vert _{AB})$ is the reduced density matrix for
lattice $A$. Obviously, the entropy of a helix state is zero since it is a
separable state. It is expected that a quenched state has {zero entropy in
the presence of }$H_{\mathrm{p}}^{1}$, while has increasing entropy in the
presence of $H_{\mathrm{p}}^{2,3}$. In addition, the sum of the spin
current at a given node $k$ is defined as%
\begin{equation}
J_{k}(t)\sum_{j=1}^{n}\left\langle \Psi (t)\right\vert \hat{\jmath}%
_{kj}\left\vert \Psi (t)\right\rangle.  \label{Jk}
\end{equation}%
{It is expected that for a quenched state we have }$J_{k}(t)=0${\ in the
presence of }$H_{\mathrm{p}}^{1}${, while }$J_{k}${\ is time-dependent in
the presence of }$H_{\mathrm{p}}^{2,3}${.}

In Fig. \ref{fig3}, we plot the entanglement entropy and current
obtained from systems with different perturbation terms. It indicates that
the result accords with our prediction. Both analytic analysis and numerical
simulation show that perturbation $H_{\mathrm{p}}^{2,3}$ breaks the static
balance, while $H_{\mathrm{p}}^{1}$ maintains the nature of a helix state.

\section{Summary}

\label{Summary}

In summary, we have studied the possible extension of a constructing phantom
helix state in a uniform system to those with irregular geometries by the
BBM. The underlying mechanism is based on the theorem about product states.
A summation of a set of sub-Hamiltonians shares a common eigenstate of each
one, if it is a spin helix state. This allows us to construct a system
possessing phantom helix states by two types of Heisenberg XXZ dimers. This
method is applicable to  quantum lattice systems with other degrees of
freedom. As a demonstration of this method, we construct a set of Hermitian
systems and study the properties of the helix eigenstates. We find that the
corresponding spin currents and helical phases obey the  analogs of
Kirchhoff's laws. In addition, the dynamic response of the helix states\ for
three types of perturbations is also investigated analytically and
numerically. Our findings provide a way to study quantum spin systems with
irregular geometries beyond the Bethe ansatz approach.

This work was supported by the National Natural Science Foundation of China
(under Grant No. 12374461).

\end{document}